# Ultra-large Mode Area Microstructured Core Chalcogenide Fiber Design for Mid-IR Beam Delivery

Ajanta Barh,[1] Somnath Ghosh,[2] R. K. Varshney,[1] and Bishnu P. Pal[1]

[1]Department of Physics, Indian Institute of Technology Delhi, New Delhi 110016, India
[2]Institute of Radio Physics and Electronics, University of Calcutta, Kolkata 700009, India

**Abstract:** An all-solid large mode-area (LMA) chalcogenide-based microstructured core optical fiber (MCOF) is designed and proposed for high power handling in the mid-IR spectral regime, covering the entire second transparency window of the atmosphere (3 – 5 μm). The core of the proposed specialty fiber is composed of a few rings of high index rods arranged in a pattern of hexagon. Dependence of effective mode area on the pitch and radius of high index rods are studied. Ultra-high effective mode area up to 75,000 μm$^2$ can be achieved over this specific wavelength range while retaining its single-mode characteristic. A negligible confinement loss along with a low dispersion slope (~ 0.03 ps/km-nm$^2$) and a good beam quality factor ($M^2$ ~ 1.17) should make this LMA fiber design attractive for fabrication as a potential candidate suitable for high power, passive applications at the mid-IR wavelength regime.

## 1. Introduction

The mid-infrared (mid-IR) wavelengths (~ 2 - 10 μm) have recently become increasingly important due to their potential applications in areas as wide as astronomy, climatology, civil, medical surgery, military, biological spectroscopy, semiconductor processing, optical frequency metrology, optical tomography and sensing [1–5]. This has opened up a wide interest in the development of optical fibers that can efficiently generate mid-IR light [6-10] and passive fibers for high power delivery [11-14]. LMA fibers are very attractive for guiding and delivering high laser power because of enhanced threshold power limit for material damages to occur. Additionally, its relatively wider core significantly reduces fiber non-linear effects like Stimulated Brillouin Scattering (SBS) inside the fiber along its length. Moreover, low numerical aperture (NA) LMA fibers are very effective in reducing the amplified spontaneous emission (ASE). On the other hand, as an active medium, LMA fibers are widely used to amplify intense pulses of single frequency signals [15]. It is well known that the overall dispersion behavior of such LMA fibers is mainly dictated by the nature of the chosen materials. Thus, for any design of mid-IR LMA fibers, fiber materials should be very carefully chosen. Material should be chemically stable, sufficiently transparent in the desired spectral range, possess low propagation loss, and drawable in fiber form. Chalcogenide glasses (S-Se-Te based glass compositions) match almost all the above-mentioned criteria, and hence are strong candidates for realizing mid-IR optical fibers. In addition, their chemical durability, glass transition temperature ($T_g$), strength, stability etc can be modified appropriately by doping with As, Ge, Sb or Ga, thereby making them suitable for drawing into an optical fiber. Though expensive their state-of-the-art fabrication technology is also well matured [5, 16-19] for realizing application-specific specialty fibers for the mid-IR wavelengths.

Various attempts have already been made to realize LMA fibers based on higher order mode (HOM) guiding fibers, in which large differential loss is induced between the fundamental mode (FM) and the HOMs. For example, in a leaky channel fiber, or hollow and solid core MOFs, or in Bragg fibers, or differential gain guiding in multi mode fibers – in all of which differential confinement loss/bend-loss can be exploited to realize effective FM operation [11-14, 20-22]. Most of these already reported design philosophies revolve around specifically targeted applications. In the context of LMA fiber designs, discrete photonic structures have recently opened up a novel design route to fulfill the task of LMA with better design freedom in controlling the optical properties of the mode and for further up-scaling of the mode size [23-24]. Our present fiber design consists of a microstructed core in place of a uniform refractive index core.

In this paper, our primary goal is to design a chalcogenide glass-based, microstructured-core LMA fiber for high power delivery at the 3 – 5 μm wavelength regime. Single-mode operation over this targeted spectral range is confirmed by maintaining the effective $V$ number of the suitably structured core within the single-mode limit. Dependence of effective mode-area ($A_{eff}$) on various fiber parameters has been studied to achieve a $A_{eff}$ as high as 75,000 μm$^2$ while maintaining very low confinement loss and a good beam quality factor ($M^2$ ~ 1.17). In our proposed fiber design route to optimize the low loss, ultra large mode area fiber with the above-mentioned good beam quality, we have essentially synthesized two different categories of optical systems: a conventional TIR guided fiber structure and a discrete photonic structure.

## 2. Proposed fiber design and numerical simulations

Mode effective area $A_{eff}$ of a fiber can be increased either by increasing the core diameter ($d$) or by decreasing the NA. Increase in $d$ makes the fiber multimoded; on the other hand due to limitations in rare earth (RE) doping level, NA in conventional fibers is hard to reduce to sufficiently small values for this purpose. In our design, we could significantly reduce the NA through an effective decrease in core refractive index ($n_c$). Instead of using a large uniform core with conventional microstructured cladding, we considered a microstructured core embedded in a material, which also formed the cladding of uniform low index. The cross section of the proposed fiber design is shown in Fig. 1. The microstructured core of the fiber consists of 4 rings of hexagonally arranged high index circular rods of chalcogenide glass $As_{20}Se_{80}$, whose refractive index ($n_r$) at 4 μm wavelength is ~ 2.575. The radius of each high index rod is denoted as $r$ and the center to center spacing between two nearest rods i.e. the pitch is denoted as $\Lambda$. Cladding forms the uniform low index background made out of chalcogenide glass composition, $Ge_{12.5}As_{20}Se_{67.5}$, whose refractive index ($n_b$) at 4 μm is ~ 2.565. Thus the core, consisting of these two chalcogenide glasses, effectively reduces the $n_c$ below fabrication limit and hence provide an ultra-high modal area.

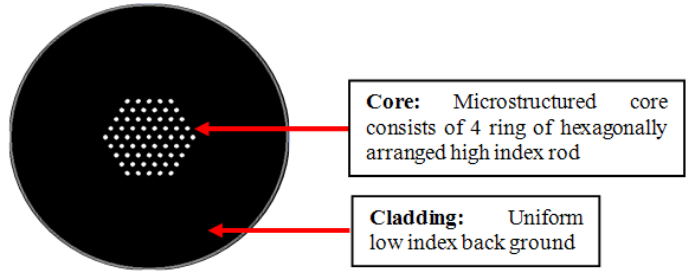

**Fig. 1.** Transverse view of the proposed LMA fiber. Microstructured core (white circles) consisting of 4 rings of hexagonally arranged $As_{20}Se_{80}$ rods are embedded at the center in $Ge_{12.5}As_{20}Se_{67.5}$ (shown in black), which also forms the cladding.

The simulation results were obtained through use of the freely available software CUDOS® based on Multipole Method, which were also verified by full vectorial finite element method. The structure was optimized with 4 rings of high index rods for SM operation. The mode effective area was calculated by using the following standard formula:

$$A_{eff} = \frac{\left(\int EE^* dA\right)^2}{\int \left(EE^*\right)^2 dA} \quad (1)$$

where $E$ and $E^*$ represents the electric field and its complex conjugate, respectively.

Dependence of $A_{eff}$ of the fundamental mode (FM) on the microstructured core parameters $r$ and $\Lambda$ is studied and the same is shown in Fig. 2. From this figure it is evident that $A_{eff}$ gradually increases with increase in $\Lambda$, while it decreases with larger values of $r$. Wave guidance in this geometry can be physically understood as follows:

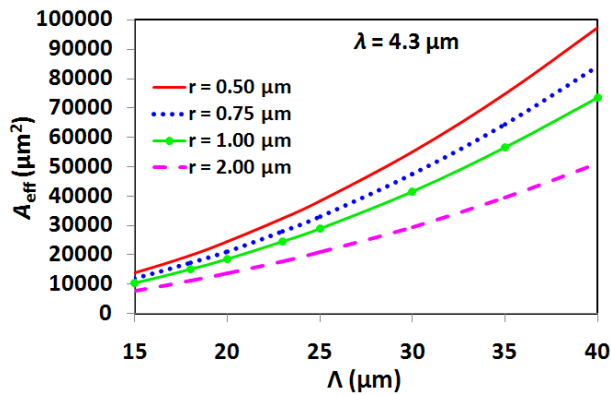

**Fig. 2.** Variation of mode effective area of FM with pitch ($\Lambda$) for 4 different values of $r$ as the labeling parameter.

In the proposed geometry with optimum fiber structure, the supermode of all the modes individually guided by each of the high index rods essentially forms the FM when they get phase matched at the operating wavelength. Increase in $\Lambda$ here implies an effectively larger core boundary and hence larger mode area. On the other hand, as $r$ decreases, the individual mode field spreads out more in low index region, overlap between these modes increases and hence the overall size of supermode gets larger. All these features lead to an effective increase in $A_{eff}$ of the fiber so much so that with appropriate optimization of the core parameters for this pair of chalcogenides, we could theoretically demonstrate $A_{eff}$ as high as 75,000 $\mu m^2$ for the FM.

The modal field distributions corresponding to two different values of pitch ($\Lambda$ = 20 $\mu$m and 35 $\mu$m) and three values of $r$ ($r$ = 1.5 $\mu$m, 1.0 $\mu$m and 0.5 $\mu$m) are shown in Fig. 3 (a) – 3(f). From these field plots, it can be seen that for larger values of $\Lambda$, the discrete nature of the system governs the propagation of light. In other words, overlap among the individual eigen modes decreases and larger amount of fractional field powers gets confined in the high index rod region. As a result, the overall field distribution appears to be more discrete in nature. However, for smaller values of $r$, the eigen modes increases and hence supermode area increases resulting in the overall field distribution to be more uniform.

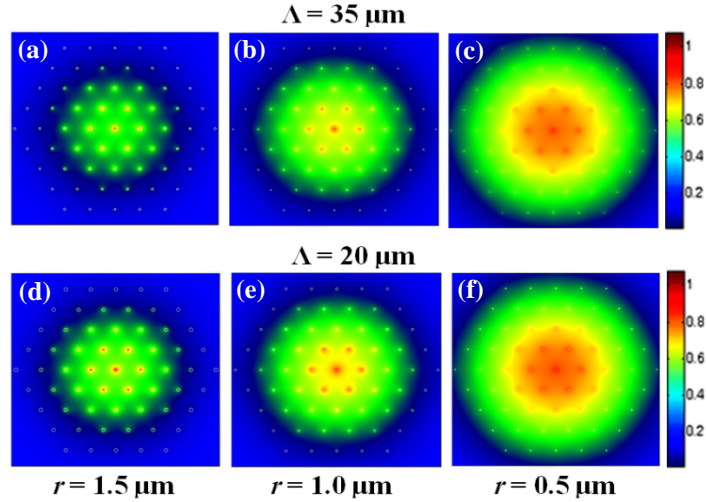

**Fig. 3.** Fundamental mode pattern at $\lambda$ = 4.3 $\mu$m. (a) $r$ = 1.5 $\mu$m and $\Lambda$ = 35 $\mu$m; (b) $r$ = 1.0 $\mu$m and $\Lambda$ = 35 $\mu$m; (c) for $r$ = 0.5 $\mu$m and $\Lambda$ = 35 $\mu$m; (d) $r$ = 1.5 $\mu$m and $\Lambda$ = 20 $\mu$m; (e) $r$ = 1.0 $\mu$m and $\Lambda$ = 20 $\mu$m; (f) for $r$ = 0.5 $\mu$m and $\Lambda$ = 20 $\mu$m.

In parallel, to confirm the single mode operating condition, the fiber $V$ number is calculated by approximating this MOF structure to an equivalent step index fiber (E-SIF) structure. In general for E-SIF, $V$ number is defined through [25]

$$V = \frac{2\pi a}{\lambda}\sqrt{n_c^2 - n_b^2} \qquad (2)$$

where $\lambda$ is the operating wavelength, $a$ is the core radius, $n_c$ is the effective core index and $n_b$ is the cladding index. Moreover, as there is no well defined boundary between core and cladding region, we have approximated $a$ to be 4 x $\Lambda$ as 4 rings of rods are considered following usual practice. Refractive index $n_b$ is same as the background index and $n_c$ is calculated by averaging the refractive index over the core region [24], where $n_c$ is taken as

$$n_c = \sqrt{\frac{A_l n_b^2 + A_h n_r^2}{A_t}} \qquad (3)$$

where $A_l$ and $A_h$ are the respective areas of low and high refractive index regions inside the core, and $A_t$ is the total core area. Figure 4 shows the variation of $V$ with wavelength for a chosen fixed pitch ($\Lambda$ = 35 $\mu$m) and different values of $r$. It is clearly evident from the figure that, for $\Lambda$ = 35 $\mu$m, $V$ number remains below 2.4 [25] up to $r \approx 0.75$ $\mu$m for the desired wavelength ($\lambda$) range and hence the proposed fiber is effectively operates as a single-mode fiber within the mid-IR regime covering 3 ~ 5 $\mu$m.

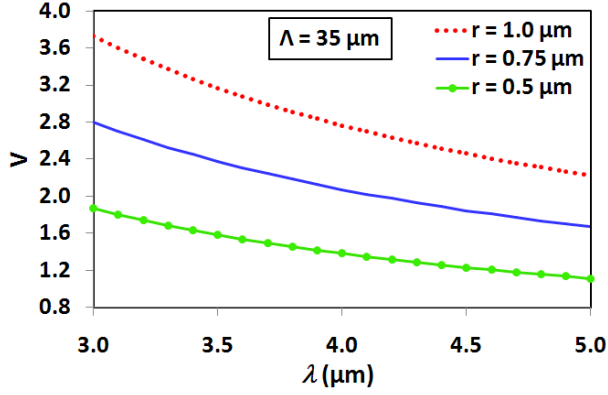

**Fig. 4.** Variation of $V$ number of E-SIF with the operating wavelength for $\Lambda = 35$ μm and $r = 0.5$ μm, 0.75 μm and 1.0 μm.

The variation of $A_{eff}$ with wavelength ($\lambda$) is also plotted for $\Lambda = 35$ μm and $r = 0.5$ μm, 0.75 μm and shown in Fig. 5. The gradual increment of modal area with $\lambda$ implies that the fundamental supermode is a real guided mode of the fiber. It can be further noted that, for a chosen value of $\Lambda$, the value of $A_{eff}$ is almost 20,000 μm² higher for $r = 0.5$ μm in comparison to its counterpart at $r = 0.75$ μm. This figure also confirms that for $r = 0.5$ μm, $A_{eff}$ as high as 75,000 μm² can be achieved near an operating wavelength (~ 4.5 μm).

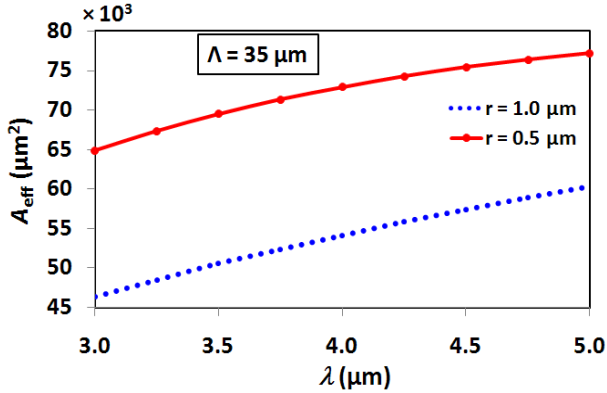

**Fig. 5.** Variation of $A_{eff}$ with $\lambda$ for $\Lambda = 35$ μm and $r = 0.5$ μm, 0.75 μm.

We have also studied the dispersion characteristics of the designed LMA fiber. Variation of the dispersion coefficient ($D$) and dispersion slope ($S_D$) with $\lambda$ is shown in Figs. 6(a) and 6(b), respectively. From Fig. 6(a) it is clear that the structure possesses normal dispersion over the entire wavelength range. Dispersion slope is very small (cf. Fig. 6(b)) and lies between 0.01 – 0.07 ps/km.nm² over the wavelength range 3 ~ 5 μm. With this low dispersion values, our design would enable almost distortion free high power light propagation in the form of FM.

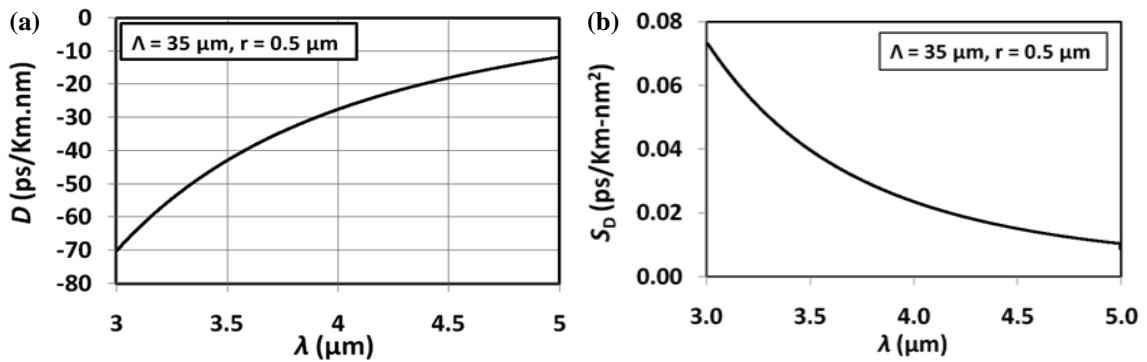

**Fig. 6.** (a) Dispersion coefficient ($D$) vs wavelength for the designed ultra-LMA fiber; (b) Dispersion slope ($S_D$) vs wavelength. Both are plotted for $\Lambda = 35$ μm and $r = 0.5$ μm.

We have also studied the key features of the proposed LMA design like the beam quality factor ($M^2$) to check its single modeness and corresponding confinement loss for its potential applications. For a fiber design with a chosen set of parameters of $\Lambda = 35$ μm and $r = 0.5$ μm, the $A_{eff}$ is calculated to be ≈ 75,000 μm$^2$, with a negligible confinement loss ( ≈ $10^{-10}$ dB/m) near $\lambda = 4$ μm and the corresponding values of $D ≈ -20.6$ ps/km.nm, $S_D ≈ 0.03$ ps/km.nm$^2$ and $M^2 ≈ 1.17$ at $\lambda = 4.0$ μm. As the calculated $M^2$ value is very close to 1.1, beam quality at the output of the designed fiber length should be excellent.

For this particular set of design parameters, the $z$-component of the pointing vector ($S_z$) is plotted in Figs. 7(a) and 7(b). The kinks that appear at 5 locations along the direction $y$ (cf. Fig. 7(b)) essentially indicates the locations of high index rods along $y$ direction at $x = 0$ and is the results of higher power concentration at those regions.

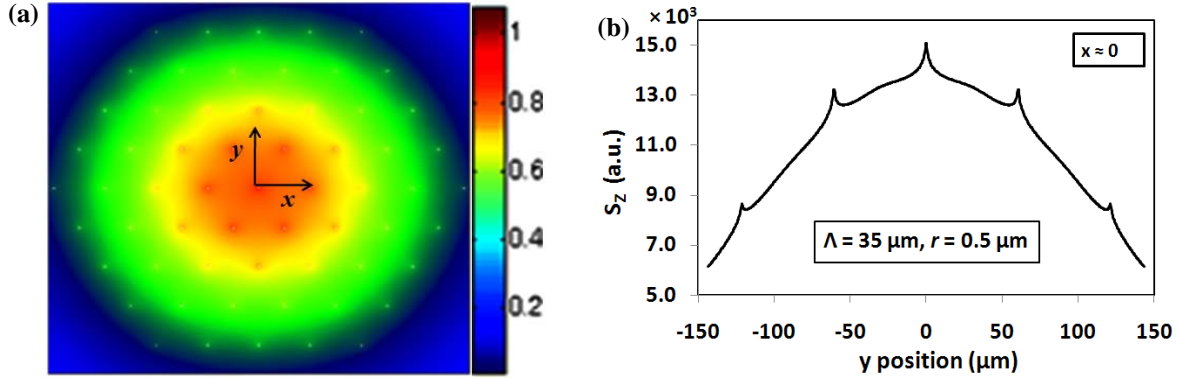

**Fig. 7.** Pointing vector ($S_z$) plot for optimum structure, $\Lambda = 35$ μm and $r = 0.5$ μm. (a) $x - y$ plot of $S_Z$; (b) $y$ variation at $x = 0$ position.

Then we also searched for the next higher order mode (HOM) in order to find exact single-mode operating wavelength regime. We have found that for $r = 0.5$ μm, no HOM exists. However, for $r = 0.75$ μm and $\Lambda = 35$ μm, first HOM exists at $\lambda < 4$ μm, which is equivalent to LP$_{11}$ mode [cf. Figs. 8(a), 8(b) and 8(c)]. The confinement loss of this mode is as high as ~ 13 dB/m. We may mention that some salient features of these results were reported by us at the international symposia, JSAP-OSA Joint Symposia 2012 held at Matsuyama, Japan.

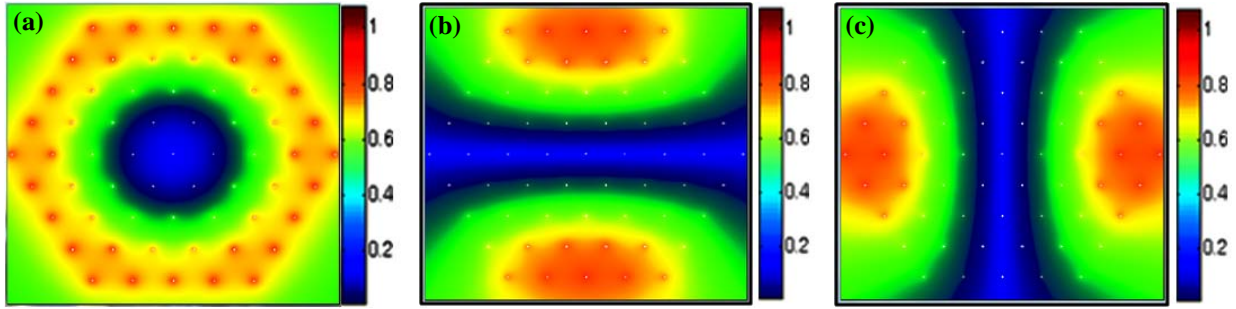

**Fig. 8.** First HOM for $r = 0.75$ μm and $\Lambda = 35$ μm at $\lambda ≈ 4$ μm (equivalent LP$_{11}$ mode). (a) $z$ component of pointing vector ($S_z$); (b) $x$ component of electric field ($E_x$); (c) $x$ component of magnetic field ($H_x$).

## 3. Conclusions

Through a detailed numerical study, an all-solid chalcogenide-based specialty LMA fiber for mid-IR (3 ~ 5 μm) passive guidance of light has been designed. In our MOF design, instead of a solid core, a microstructured core is considered to lower the effective core refractive index and thus achieve an effectively very low NA. In a conventional fiber, it is almost impossible to go down to this level of NA through doping with refractive index modifiers. This proposed concept could be exploited with appropriate choice of material combinations (here only As$_{20}$Se$_{80}$ and Ge$_{12.5}$As$_{20}$Se$_{67.5}$ based MOF is explored) to realize ultra-large mode area fiber. We have demonstrated a theoretical $A_{eff}$ of 75,000 μm$^2$ with good beam quality ($M^2$ ~ 1.17) for single mode operation, low dispersion slope ($S_D$ ~ 0.01 - 0.07 ps/km.nm$^2$) and negligible confinement loss for entire wavelength range.

Some other issues related to LMA fiber design, like proper heat removal, bend sensitivity etc. are yet to be investigated. However for small fiber length, our methodology should serve as the initial design platform for realizing such an extremely large mode area specialty fiber matching the second atmospheric transparency window for gainful exploitation. We hope our adopted design philosophy along with the detailed analysis as presented in this paper for such large mode area designs may open a new area to synthesize three different areas of photonics namely microstructured optical fiber, discrete photonics systems and microstructured core fiber designs.

**Acknowledgement**

This work relates to Department of the Navy (USA) Grant N62909-10-1-7141 issued by Office of Naval Research Global. The United States Government has a royalty-free license throughout the world in all copyrightable material contained herein. AB gratefully acknowledges the award of a Senior PhD fellowship by CSIR (India). SG acknowledges the financial support by DST (India) as a INSPIRE Faculty Fellow [IFA-12; PH-13].